\documentstyle[preprint,prb,aps]{revtex}
\large
\begin{document}
\newcommand{\be}{\begin{equation}}
\newcommand{\ee}{\end{equation}}
\newcommand{\bea}{\begin{eqnarray}}
\newcommand{\eea}{\end{eqnarray}}
\draft
%\twocolumn[\hsize\textwidth\columnwidth\hsize\csname @twocolumnfalse\endcsname

\title{
Interacting one dimensional electron gas with open boundaries
}
\author{M. Fabrizio$^{(a,b)}$ and Alexander O. Gogolin$^{(a,c)}$
}
\address{
$^{(a)}$Institut Laue--Langevin, B.P.156 38042 Grenoble, Cedex 9, France\\
$^{(b)}$Institut for Advanced Studies, Via Beirut 4, 34014 Trieste, Italy\\
$^{(c)}$Landau Institute for Theoretical Physics, Kosygina str. 2,
Moscow, 117940 Russia
}
\date{\today}
\maketitle
\begin{abstract}
We discuss the properties of interacting electrons on a finite
chain with open boundary conditions.
We extend the Haldane Luttinger liquid description to these systems
and study how the presence of the boundaries modifies various correlation
functions. In view of possible experimental applications to quantum wires,
we analyse how tunneling measurements can reveal the underlying
Luttinger liquid properties. The two terminal conductance is calculated.
We also point out possible applications to quasi one dimensional
materials and study the effects of magnetic impurities.
\end{abstract}
%\pacs{ }
%]
\narrowtext

%\newpage

\section{Introduction}
\label{sec:Introduction}

Physical properties of one dimensional (1D) metals are by now well
understood theoretically. Unlike higher dimensional metal, where
interaction slightly modifies the free Fermi gas behavior
(Landau-Fermi liquid picture), in 1D metals the electron-electron
interaction plays a fundamental role and strongly affects their
physical properties as compared to the ideal Fermi gas. The most interesting
feature is the absence of well defined single particle
excitations. The only stable low energy excitations turn out to be
collective charge and spin density fluctuations (zero sound modes).
Charge and spin sounds are dynamically independent, which,
together with the absence of the continuum of electron-hole
excitations, gives rise to the so-called spin-charge separation.
The interaction also modifies the asymptotic
behavior of all the correlation functions: at large distances (times)
they are shown to decay as a power law with
interaction dependent exponents (Luttinger liquid behavior, for a review
see Ref.[$\!\!$\onlinecite{1d}]).

Before the recent achievements of the submicron--size technology
in the fabrication of 1D quantum wires, 1D electron gases have been studied
either on their own rights (as interesting mathematical objects)
or in view of applications to quasi 1D materials\cite{Bourbonnais}.
For these purposes,
it was sufficient to investigate infinite systems (or to impose periodic
boundary conditions, which are relatively easy to treat).
Semiconductor devices with 1D confining potentials (quantum wires)
represent a new and promising realizations of 1D electron systems
which provide an alternative way to study experimentally
the Luttinger liquid properties. For instance, the collective nature
of the low energy excitations have been successfully probed by
inelastic light scattering experiments\cite{Goni}.
Other experiments particularly suited for quantum wires are those measuring
transport properties, and an interesting question is
whether these
measurements are able to reveal the Luttinger liquid character of 1D systems.
As regards to the measurements of bulk properties in a clean system
(e.g. optical conductivity) the answer would be no. In fact,
although many properties are anomalous,
the (bulk) transport properties of a clean
1D metal are expected to be qualitatively similar to those of an ordinary
3D metal.
As regards to the surface measurements or, more generally, to the response to
any local probe,
it is claimed that the answer is positive. This result has been reached by
extending
the analysis of idealized 1D chains
to physical quantum wires which are finite systems with open boundaries
or contacts of 1D electron gases with normal 3D metals.

The anomalous response of an interacting 1D Fermi gas to a local probe
was firstly recognized by Kane and Fisher\cite{K&F}.
Subsequently many different experimental situations have been
proposed and analysed in the framework of the Luttinger liquid theory of
1D metals\cite{everythingelse}. Several effects due to open boundary
conditions have been studied by means of conformal field theory\cite{Aff}.
All these studies suggest that the interaction affects (in a non trivial way)
the behavior of the Fermi gas close to the boundaries. Consequently
transport measurements probing the edge
properties should in principle reveal the Luttinger liquid behavior.

In this paper we study the properties
of a 1D chain of interacting electrons with open boundary conditions
from a more traditional point of view not assuming
conformal invariance. Some of the results will be new,
some not; our intention is to provide a simple
but powerful tool to tackle different problems that
arise while studying finite 1D systems.
%We will show how different
%correlation functions are modified by the presence of the boundary,
%and how this modification will affect some transport measurement,
%e.g. a two terminal conductance.

The layout of the paper is as follows.
In Section II we develop the
bosonization technique appropriate for open boundary conditions,
generalizing Haldane's approach (which was originally deviced
for periodic systems). The bosonization for non interacting
electrons is discussed first. Then we study the interacting case.
In Section III several correlation
functions are calculated and the Friedel oscillation caused by the
boundary is discussed. We also give an alternative derivation,
based on Ward's identity method, of correlation functions and outline
some applications of the open boundary analysis to slightly doped
quasi-1D electron systems.
Section IV is devoted to the analysis of
transport phenomena in quantum wires.
The two terminal conductance is calculated and some other tunneling processes
at the edge of the wire are discussed.
In Section V we study
a magnetic impurity coupled to a 1D electron gas in the presence of a
potential scattering.
There are four appendices. In Appendix A some formulae used in computation of
correlation functions are given.
The
remaining appendices are devoted to extensions of our approach:
an X-ray edge type problem related to the scattering potential sign changing
operator
is investigated in Appendix B, effects of long-range electron-electron
interaction are examined in
Appendix C, and gapped phases of finite systems are studied in Appendix D.

\section{Open boundary bosonization }
\label{sec:Bosonization}

This section is intended to give a general method of
bosonizing an interacting Fermi--system in the case of
open boundaries (i.e. of a finite system of the length $L$).

The bosonization method has a long history.
The equivalence between the excitation spectra of interacting
fermions and free bosons in one dimension (1D) was established
by Mattis and Lieb \cite{ML} in their solution of the Luttinger model
\cite{Lutt}. The bosons were identified as particle--hole
excitations over the Fermi sea, and their dynamics turned out to
be that one predicted by random--phase approximation.
However, the full power of the bosonization method
became clear later, when the representation for the electron creation
(annihilation) operator in terms of free bosons was discovered
\cite{LP,Matt,Mand}. This provided a powerful tool to calculate
fermionic correlation functions in terms of free boson correlation
functions. (Actually, this representation is very close in spirit to the
famous Jordan and Wigner representation.)
The finite size effects (assuming periodic boundary conditions)
were studied by Haldane \cite{Haldane}.

We start with the Hamiltonian:
\[
H=H_0+H_{int}\;,
\]
where the first term represents the kinetic energy,
\be
H_0= \sum_{s=\uparrow ,\downarrow}\int_{0}^{L} dx
\psi^{\dagger}_{s} (x)\varepsilon (-i\partial_x )
\psi^{\phantom{\dagger}}_{s} (x)\; ,
\label{ke}
\ee
and the second one
describes the electron-electron interaction,
\be
H_{int}=\frac{1}{2}\sum_{ss'}\int dxdy
\psi^{\dagger}_{s}(x)\psi^{\dagger}_{s'}(y)U_{ss'}(x-y)
\psi^{\phantom{\dagger}}_{s'}(y)\psi^{\phantom{\dagger}}_{s}(x) \; ,
\label{int}
\ee
$\varepsilon (k)$ is the dispertion law of the
1D band, and $\psi_s (x)$ is the spin $s$ electron
annihilation operator subject to the open boundary conditions:
\be
\psi_s (0)=\psi_s (L)=0
\label{boundcond}
\ee

\subsection{Non-interacting electrons}

We consider the non-interacting case first.
This situation has already been discussed in the literature
(see e.g. Ref.[$\!\!$\onlinecite{Ludwig}]),
but we still find it useful to give
all the details which we are going to use while studying the
interacting case.

The Fourier mode expansion of the $\psi$-operator, appropriate
for the boundary conditions (\ref{boundcond}), takes the form:
\be
\psi_s (x)=\sqrt{\frac{2}{L}}\sum_k \sin (kx)c_{sk}
\label{standing}
\ee
with $k=\pi n/L$, $n$ being a positive integer. The single
electron spectrum is $\varepsilon (k)$ and the Fermi surface
consists of the {\em single} point $k=k_F$ (see Fig.\ref{fermi}).
Concentrating on
the vicinity of this point we define slow varying right and
left moving fields:
\be
\begin{array}{l}
\psi_{s R} (x)=-\frac{\displaystyle i}{\displaystyle\sqrt{2L}}\sum_{p}
{\rm e}^{ipx}c_{s,k_F+p} \; ,\\
\psi_{s L} (x)=\frac{\displaystyle i}{\displaystyle\sqrt{2L}}\sum_{p}
{\rm e}^{-ipx}c_{s,k_F+p}
\end{array}
\label{Rff}
\ee
such that ($p=\pi n/L$)
\be
\psi_s (x)= {\rm e}^{ik_F x} \psi_{sR} (x) +
 {\rm e}^{-ik_F x} \psi_{sL} (x) \;.
\label{decomposition}
\ee
These fields, however, are {\em not} independent, as in the case
of periodic boundary conditions \cite{Haldane}, but satisfy:
\be
\psi_{s L} (x)= - \psi_{s R} (-x)\; .
\label{RL}
\ee

Therefore, one can actually work with the right moving field
only, the left moving one is then defined by the above relation.
The boundary condition
\[
\psi_s (0)=0
\]
is automatically satisfied, whereas the condition
\[
\psi_s (L)=0
\]
implies that the operator $\psi_{s R} (x)$ should obey
\[
\psi_{s R} (-L) =\psi_{s R} (L) \; .
\]
So we can regard the field $\psi_{s R} (x)$ as defined
for all $x$ but obeying the periodicity condition with the
period $2L$:
\be
\psi_{s R} (x+2L) =\psi_{s R} (x) \; .
\label{perstress}
\ee

In terms of the right moving field, the kinetic energy
(\ref{ke}) takes the form:
\[
H_0= v_F \sum_{s} \int_{-L}^{L} dx
\psi^{\dagger}_{s R} (x)(-i\partial_x )
\psi^{\phantom{\dagger}}_{s R} (x) \;
\]
where we have linearized the electron spectrum and
the energy is accounted for from the Fermi energy of a
reference system with $N_{s 0}$ number of spin $s$ electrons.

The right moving Fermi field (\ref{Rff}) obeys {\em periodic}
boundary condition (\ref{perstress}), so it can
straightforwardly be bosonized. We will
conveniently employ the following version of the
bosonization formula:
\be
\psi_{s R} (x) = \frac{\eta_s}{\sqrt{2L}}
{\rm e}^{- i\theta_s}
{\rm e}^{ i \pi \frac{x}{L} \Delta N_s }
{\rm e}^{i \phi_s (x)} \; ,
\label{fermiop}
\ee
where $\Delta N_s$ is the number of extra electrons with spin $s$,
\[
\Delta N_s = N_s-N_{s 0} \; ,
\]
the variable $\theta_s$,
canonically conjugate to $\Delta N_s$,
\[
\left[ \theta_s, \Delta N_s \right] =i \; ,
\]
is defined modulo $2\pi$. The operators $\eta_s$ are real (Majorana) fermions,
\[
\{\eta_\uparrow ,\eta_\downarrow\}=0\;,\eta_s^2=1\;,
\]
which stand to assure the correct anticommutation
rules for electron operators with different spin $s$.
The phase field $\phi_s (x)$ is given by
the expression
\[
\phi_s (x) = \sum_{q>0} \sqrt{\displaystyle \frac{\pi}{q L}}
{\rm e}^{iqx -\alpha q/2} b_{q} + {\rm H.c.}
\]
and satisfies periodic boundary condition:
\be
\phi_s (x+2L) = \phi_s (x) \; .
\label{phi-period}
\ee
Here $b_q$ are canonical Bose operators;
$q=\pi n/L$, $n$ is an integer, and
we have introduced a high--energy cut--off $\alpha$.
It is straightforward to check that the operators $\psi_{sR}(x)$,
defined by Eq.(\ref{fermiop}), obey standard fermionic commutation
relations (in the limit $\alpha \to 0$).

Alternatively one could write the Fermi operator (\ref{fermiop})
in a normal ordered form, noticing:
\[
\frac{1}{\sqrt{2L}}:{\rm e}^{i\phi_s (x)}: \to
\frac{1}{\sqrt{2\pi\alpha}}{\rm e}^{i\phi_s (x)} \; ;
\]
we shall use both normal ordered and not normal ordered forms
of Fermi operators.

The presence
of the momentum space cut-off $\alpha$
reflects the finite band--width of the original electron band
($\alpha$ is understood to be much larger than the
lattice spacing; actually, $1/\alpha$ is the region around
$k_F$ where the electron band spectrum can be linearized).

The density of right moving electrons is given by
\[
\rho_{s R} (x) = \frac{\Delta N_s }{2L} + \frac{\partial_x
\phi_s (x)}{2\pi} \; .
\]
We notice that
\[
\rho_{s L} (-x) =\rho_{s R} (x)\; .
\]
The bosonized form of the kinetic energy is:
\[
H_0=\pi v_F \sum_{s}\int_{-L}^{L} dx
:\rho_{s R} (x)\rho_{s R} (x): =
v_F\sum_{s q>0}
q b^\dagger_{s q}b^{\phantom{\dagger}}_{s q}
+\frac{\pi v_F}{2L}(\Delta N_s )^2  \; .
\]

Before we turn to the interaction effects, we define the
bosonic variables corresponding to charge and spin excitations:
\[
b_{\rho(\sigma) q}=\frac{1}{\sqrt{2}}\left(b_{\uparrow q}\pm
b_{\downarrow q}\right)\;,
\]
and
\[
\Delta N_{\rho(\sigma)}=\Delta N_{\uparrow} \pm \Delta N_{\downarrow}\;.
\]

\subsection{Interaction effects}

In order to make use of the above bosonization procedure also for the case
of interacting electrons, we try the {\em same} trick; namely, we express
the part of the Hamiltonian, which is responsible for
the interaction, Eq.(\ref{int}), in terms of the right moving Fermi field
$\psi_{s R}$ only. The cost is that the resulting expression is
highly non-local in space as illustrated in Fig.\ref{right}.
Nevertheless, as we show below,
at least in the case of short--range electron-electron interaction
$U(x-y)$, the problem can be quite simply treated in terms of bosonic fields.

Here we focus on the case of $R \simeq \alpha$, where $R$ is the characteristic
radius of the region where the function $U(x-y)$ is essentially non--zero.
Then the $q$-dependence of corresponding Fourier transforms can conveniently
be neglected. (This implies that all the distances considered should be
much larger than $R$.)

The interaction part of the Hamiltonian contains several terms
classified in what follows.

The terms, diagonal in the electron densities, can be written in
conventional way:
\begin{equation}
\frac{g_{\rho(\sigma)}}{2}\int_{0}^{L}dx \left[
\rho_{\rho(\sigma) R} (x)\rho_{\rho(\sigma) R} (x)
+\rho_{\rho(\sigma) L} (x)\rho_{\rho(\sigma) L} (x)\right]=
\frac{g_{\rho(\sigma)}}{2}\int_{-L}^{L}dx
\rho_{\rho(\sigma) R} (x)\rho_{\rho(\sigma) R} (x) \;.
\label{gint}
\end{equation}
This renormalizes the sound velocity:
\[
v_{\rho(\sigma)}^0= v_F + \frac{g_{\rho(\sigma)}}{2\pi}\;.
\]

The term mixing right and left densities is of the form:
\be
\tilde{g}_{\rho(\sigma)}\int_{0}^{L}dx
\rho_{\rho(\sigma) R} (x)\rho_{\rho(\sigma) L} (x)=
\frac{1}{2}\tilde{g}_{\rho(\sigma)}\int_{-L}^{L}dx
\rho_{\rho(\sigma) R} (x)\rho_{\rho(\sigma) R} (-x)\;,
\label{gtildeint}
\ee
i.e. it is non-local in space (see Fig.\ref{right}). Still, this term is
quadratic in the electron densities and
therefore takes a simple form in terms of bosonic operators
(the remaining terms, which do not assume a form quadratic in densities
are discussed in the last part of this section).

Consequently, the Hamiltonian becomes:
\be
H=\sum_{\nu=\rho(\sigma)}\left\{ \sum_{q>0} v_\nu^0
q \left[ b^\dagger_{\nu q}b^{\phantom{\dagger}}_{\nu  q}
-\frac{\tilde{g}_{\nu}}{4\pi}\left(
b^{\phantom{\dagger}}_{\nu  q}b^{\phantom{\dagger}}_{\nu  q}+\
b^\dagger_{\nu q}b^\dagger_{\nu q}\right)\right]
+ \frac{\pi v_{\nu N}}{4 L} (\Delta N_\nu )^2 \right\} \; ,
\label{ham-nondiag}
\ee
where
\[
v_{\nu N}= v_F + \frac{g_\nu + \tilde{g}_\nu}{2\pi}.
\]
This can be diagonalized in a standard way by the Bogolubov rotation
\be
b_{\nu q}\to \cosh(\varphi_{\nu}) b_{\nu q} -
\sinh(\varphi_{\nu}) b^\dagger_{\nu q},
\label{rotation}
\ee
where
\be
\tanh(2\varphi_{\nu})=-\frac{\tilde{g}_{\nu}}{2\pi v_\nu^0}\;.
\label{defphi}
\ee
[Notice that whereas the rotation angles $\varphi_\nu$ are defined in the
same way as in the bulk case, Eq.(\ref{defphi}), there is an important
difference in sign in Eq.(\ref{rotation}).]
This rotation is achieved by the canonical transformation
\be
H\to UHU^\dagger =\sum_\nu \left\{ \sum_{q>0} v_\nu
q  b^\dagger_{\nu q}b^{\phantom{\dagger}}_{\nu  q}
+ \frac{\pi v_{\nu N}}{4 L} (\Delta N_\nu )^2 \right\} \; ,
\label{canonical}
\ee
where
\[
v_\nu=\frac{v_\nu^0}{\cosh(2\varphi_{\nu})}\;.
\]
The previously defined $v_{\nu N}$ can be alternatively written as
\begin{equation}
v_{\nu N}=\frac{v_\nu}{K_\nu},
\label{vnuN}
\end{equation}
being
\[
K_\nu=\exp (2\varphi_\nu)\;.
\]
The unitary operator $U$ is defined by:
\be
U=\exp\left\{\frac{1}{2}\sum_{\nu,\;q>0}\varphi_{\nu}
\left(b^\dagger_{\nu q} b^\dagger_{\nu q} -
b^{\phantom{\dagger}}_{\nu  q}b^{\phantom{\dagger}}_{\nu  q}
\right)\right\}
\label{unitaryop}
\ee

The next step is to find how the Fermi operators transform by applying $U$.
Employing the method of Lieb and Mattis\cite{ML},
after lengthy but straightforward calculations
we arrive at the main result of this section,
i.e. the expression for the electron annihilation operator in terms of free
bosons for the case of the interacting Fermi system
with open boundaries:
%\cite{notemultiple}:
\bea
&~&U\psi_{sR} (x,t)U^\dagger=
\frac{\eta_s}{\sqrt{2\pi\alpha}} {\rm e}^{-i\theta_s}
{\rm e}^{\frac{i\pi}{4 L} \sum_\nu v_{\nu N} t}
\label{fermiopfinal} \\
&~& \exp \left\{ i\sum_\nu \varepsilon_{\nu s}\left[
\pi\Delta N_\nu \frac{(x-v_\nu t)}{2L} +
\frac{c_\nu}{\sqrt{2}} \phi_\nu (x-v_\nu t)
-\frac{s_\nu}{\sqrt{2}} \phi_\nu (-x-v_\nu t)
\right] \right\} \nonumber
\eea
where $\varepsilon_{\nu s}$ is $+1$ unless $s=\downarrow$
and $\nu=\sigma$, when its value is $-1$.
We have defined
\[
c_\nu=\cosh(\varphi_\nu),\;\;s_\nu=\sinh(\varphi_\nu).
\]

It is important to notice that, if one would
write the expression (\ref{fermiopfinal}) in the normal ordered form,
this results in $x$-dependent pre--exponential factors:
\bea
&~&U\psi_{sR} (x,t)U^\dagger=
A_\psi \frac{\eta_s}{\sqrt{2 L}} {\rm e}^{-i\theta_s}
{\rm e}^{\frac{i\pi}{4 L} \sum_\nu v_{\nu N} t}
[P(2x)]^{\sum_\nu s_\nu c_\nu/2}
\label{fermiopfinalbis}\\
&~& : \exp \left\{ i\sum_\nu \varepsilon_{\nu s}\left[
\pi\Delta N_\nu \frac{(x-v_\nu t)}{2L}+
\frac{c_\nu}{\sqrt{2}} \phi_\nu (x-v_\nu t)
-\frac{s_\nu}{\sqrt{2}} \phi_\nu (-x-v_\nu t)
\right] \right\} :\nonumber
\eea
where the constant $A_\psi$ is given by
\[
A_\psi=\left[\frac{\pi\alpha}{L}\right]^{\frac{1}{4}\sum_\nu (K_\nu^{-1}-1)}\;,
\]
and the function $P(z)$ is defined in the Appendix A.
Although the pre--exponential factor in  (\ref{fermiopfinalbis})
is $x$-dependent, it does not
depend on $t$. This explicitates the asymmetry between space and time
coordinates in the present problem (for the translation invariant system,
the factors resulting from normal ordering are just constants \cite{Haldane}).

\subsection{Luttinger liquid picture}

We now pause to discuss how the open boundary conditions modify the standard
Luttinger liquid picture. According to the Haldane analysis\cite{Haldane},
a gapless 1D system with periodic boundary conditions is
described by a low energy Hamiltonian of the general form
\be
H = \sum_{q>0} v_S
q  b^\dagger_{ q}b^{\phantom{\dagger}}_{q}
+ \frac{\pi v_{N}}{2 L} (\Delta N)^2
+ \frac{\pi v_{J}}{2 L} J^2  \; ,
\label{canonical-Haldane}
\ee
$\Delta N$ being related to the total number
of particles, and $J$ to the total current. The different velocities
in (\ref{canonical-Haldane}) obey the relations:
\[
v_J=K v_S\,,\;\;\; v_N= \frac{v_S}{K},
\]
where $K$ is the parameter which governs the asymptotic
power low decay of all the correlation functions.
Two parameters, e.g. $v_S$ and $v_J$, are therefore sufficient
to determine all the low energy properties of the system
(concept of the Luttinger liquid universality).
For electrons with spin, due to the spin-charge separation, low energy
behavior of the system is also described by the Hamiltonian of
the form (\ref{canonical-Haldane}) in each charge and spin sector.

As we learned from the above analysis, in an interacting Fermi gas
with open boundary conditions
\begin{itemize}
\item[{\bf (i)}] spin-charge separation still occurs;
\item[{\bf (ii)}] each (spin or charge) sector is described by the Hamiltonian
(\ref{canonical}), similar to (\ref{canonical-Haldane}).
\end{itemize}
The important difference with Haldane's analysis is that the Fermi system with
open
boundaries is bosonized with the help of the right moving
Bose fields only, and therefore there is
only one (in each spin and charge sector) conserved ``topological number",
$\Delta N$. The total current $J$ is not any more conserved;
its dynamics we discuss in Appendix B.
The Luttinger liquid concept still holds since the two
parameters ($v_S$ and $v_N=v_S/K$) are sufficient to describe the low energy
behavior of the system.
In the spirit of Ref.[$\!\!$\onlinecite{Haldane}],
$v_N$ and $K$ should be treated as
{\em phenomenological} parameters, though they are related to
the interaction constants $g$ and $\tilde{g}$ which are in turn determined
by the interaction potential $U(x-y)$ in the starting
Hamiltonian (\ref{int}).

Notice that we started the above analysis with local
(in real space) interactions:
$g_\rho (x) =g_\rho \delta (x)$, etc.
(The interaction becomes non-local only {\em after}
formulating the problem in terms of right moving fields.)
If we would have readily started
with long-range interactions of radius $R\gg \alpha$
(but nevertheless $R\ll L$), the interaction term in
(\ref{ham-nondiag}) acquires off-diagonal corrections (in $q$-space).
In Appendix C
we show that these off-diagonal terms do not contribute to the
asymptotic behavior of the correlation functions for $|x|\gg R$ and
$v_F |t| \gg R$ and can therefore be neglected.
The exponents are thus determined by the
$q\to 0$ limit of the Fourier transform of
the interaction constants $g_\nu$.

As it is well known, for the case of electrons with spin,
the above considered terms -- quadratic in the electron densities --
do not account all the
possible interaction processes effective at low energy.
The remaining process is the so--called spin backscattering
process which is described by the following term in the Hamiltonian:
\be
\begin{array}{l}
\displaystyle
\frac{g_{bs}}{2} \sum_s \int_{0}^L dx \left[ \psi_{sR}^\dagger (x)
\psi_{sL}^{\phantom{\dagger}} (x)
\psi_{\bar{s}L}^\dagger (x)
\psi_{\bar{s}R}^{\phantom{\dagger}} (x) +
(R \to L)\right]\nonumber \\
\displaystyle
\mbox{ }=
\frac{g_{bs}}{2}\sum_s\int_{-L}^Ldx \psi_{sR}^\dagger (x)
\psi_{sR}^{\phantom{\dagger}} (-x)
\psi_{\bar{s}R}^\dagger (-x)
\psi_{\bar{s}R}^{\phantom{\dagger}} (x)\;,\\
\end{array}
\label{sb}
\ee
where $\bar{s}$ denotes $-s$. Using Eq.(\ref{fermiopfinal}) we find that,
under the transformation (\ref{canonical}), the spin backscattering term
takes the form:
\be
\frac{g_{bs}}{(2\pi\alpha)^2} \int_{-L}^Ldx
{\rm e}^{-i\sqrt{2K_\sigma}\phi_\sigma(x)}
{\rm e}^{i\sqrt{2K_\sigma}\phi_\sigma(-x)}
{\rm e}^{-2i K_\sigma f(2x)}\; .
\label{sbbos}
\ee
Eq.(\ref{sbbos}) can not be diagonalized in general. Nevertheless,
one can analyze the effects of the spin backscattering by
applying the renormalization group (RG) method,
as it has been done for infinite system\cite{1d}.

For $K_\sigma>1$ the operator (\ref{sbbos}) is irrelevant, and
it flows to zero under RG process, meanwhile renormalizing $K_\sigma$ to
a smaller value $K_\sigma^*\ge 1$. For spin isotropic interaction
the fixed point value is
\be
K_\sigma^* = 1\; .
\label{Kstar}
\ee
In this case, therefore, the approach we have developed above correctly
describes the
low energy properties of the system\cite{notescaling}.

On the contrary, if $K_\sigma<1$, the spin backscattering interaction is
relevant,
i.e. it flows to strong coupling under RG process. This is
interpreted as the opening of a gap in the spin excitation spectrum.
It is known that\cite{1d}, under scaling procedure,
$K_\sigma$ decreases and eventually crosses the value $K_\sigma=1/2$
at which the model has been exactly solved by Luther and Emery\cite{L&E}.
Therefore the solution at this point is believed to give a qualitatively
correct  description of the gapped phase for any $K_\sigma<1$.
The influence of open boundary conditions on the {\em gapped} phase is
analyzed in Appendix D.

Finally, we notice that although we have modeled the finite electron system
by imposing vanishing boundary conditions (\ref{boundcond}), one could
equivalently switch on a binding wall potential.
% my suggestion
%% why do not we use a V(x) which goes to V_0 for both directions, just
%% to be consistent with our starting model? Something like this:
%Namely a potential
%$V(x)$ such that $V(|x|\gg L/2)\to V_0$ but $V(|x|\ll L/2) \to 0$
%is equivalent (if $V_0$ is larger than the Fermi energy
%and if
%% I think, but I am not sure !!!
% the distance over which $V(x)$ varies is much smaller than $L$)
%to a finite system of length of order $L$.
%In this case one should use as a basis the eigenfunctions of
%this potential $\varphi_k(x)$ while defining the electron field operator:
%\be
%\psi_s(x)=\sqrt{\displaystyle\frac{2}{L}}\sum_k\varphi_k(x)c_{sp}.
%\label{eigen-standing}
%\ee
%Since at $|x|\ll L/2$
%the eigenfunctions $\varphi_k(x)$ are {\em standing waves},
%the expression (\ref{eigen-standing}) asymptotically
%coincides with (\ref{standing}) and our above considerations
%are applicable also to the binding wall potential case.
For instance a potential
$V(x)$ such that $V(x\to \infty)\to 0$ but $V(x\to -\infty) \to V_0$
is equivalent (if $V_0$ is larger than the Fermi energy) to a left boundary.
In this case one should use as a basis the eigenfunctions of
this potential $\varphi_k(x)$ while defining the electron field operator:
\be
\psi_s(x)=\sqrt{\displaystyle\frac{2}{L}}\sum_k\varphi_k(x)c_{sk}.
\label{eigen-standing}
\ee
Since at large positive $x$
the eigenfunctions $\varphi_k(x)$ are {\em standing waves},
the expression (\ref{eigen-standing}) asymptotically
coincides with (\ref{standing}) and our above considerations
are applicable also to the binding wall potential case.

\section{Correlation functions}
\label{sec:corfun}

Given the bosonized form of the electron operators,
Eqs.(\ref{fermiopfinal})-(\ref{fermiopfinalbis}),
one can compute various correlation functions. Here we discuss some examples of
physical interest.

\subsection{The Friedel oscillation}

Since the system under study is obviously not translation invariant,
the mean electron density is not homogeneous, so that the Friedel oscillation
is build in the ground state.
Namely, we obtain
\bea
\delta \rho (x)&=&
- {\rm e}^{2ik_Fx}\sum_s\langle \psi_{Rs}^\dagger(-x)\psi_{Rs}(x)\rangle
+ (x\to-x)\nonumber \\
&=& - \frac{{\rm e}^{2ik_Fx}}{\pi\alpha}
[P(2x)]^{\frac{1}{2}\sum_\nu K_\nu }\, {\rm e}^{-if(2x)}
+(x\to -x) \;.
\label{friedelosc}
\eea

Consider distances $x \ll L$. The Friedel oscillation then
takes the form:
\[
\delta \rho (x)= - \frac{1}{\pi\alpha}
\left[ \frac{\alpha}{\sqrt{\alpha^2+4x^2}} \right]^{\frac{1}{2}\sum_\nu K_\nu}
\sin (2 k_Fx) \;.
\]
This expression describes how the perturbation of the electron density,
caused by the boundaries in a semi--infinite system,
decays at large distances.

It is interesting to examine to which
extend the open boundary consideration applies to
doped systems. Kane and Fisher noticed
that a local impurity potential
\[
V(x)\sum_s \psi_s^\dagger(x)\psi_s(x),
\]
is equivalent, at low energy,
to an infinite barrier, i.e. to a boundary
(for infinite system)\cite{K&F}.
Indeed, they found that the backscattering
part of the impurity potential
flows to infinity under scaling according to:
\be
\frac{dV}{d\ln \omega}=-\frac{1}{2} \left( 1- K_\rho\right) V\;,
\label{Dsc}
\ee
where the energy $\omega$ is scaled to zero
starting from the bandwidth $\omega=D$.
Repulsive ($K_\rho<1$), spin isotropic ($K_\sigma=K_\sigma^*=1$)
interaction is assumed.
The proof is completed by studying the opposite
limit of {\em strong potential}, which results in
a {\em weak link} between two independent
semi-infinite systems - the electron
tunneling through a weak link
is described by an irrelevant operator
(which flows to zero under scaling).

Assume now that many scatterers are present in the system.
This is what happens in the case of
doped quasi-1D electron systems. Does the above argument imply that,
upon doping, the electron gas will
break up, according to the concrete
realization of the impurity potential,
into a set of almost independent segments
(to which open boundary description is applicable)
connected by weak links?
To answer this question we notice that these segments are finite,
so that there is a minimal excitation energy
(dimensional quantization), $\omega_{min}$.
Consequently, the above scaling procedure should be stopped at energies
$\omega \sim \omega_{min}$. The relevant question
therefore is whether $\omega_{min}$ is so small
that the impurity potential grows large enough to
produce effectively infinite barriers
{\em before} $\omega$ reaches $\omega_{min}$. Obviously,
this is just a condition on the
scatterers concentration.
% my suggestion
The energy $\omega_{min}$
should be compared with the cross over scale,
$\omega_0$, between the
weak coupling regime of the impurity potential and
the strong coupling one (the latter being equivalent to
a regime of tunneling trough a weak link).
The latter scale can be estimated by imposing
the solution of Eq.(\ref{Dsc}) to be of the order
of the bandwidth. That gives:
\[
\omega_0\sim D \left(\frac{V_0}{D}\right)^{\displaystyle
2/(1-K_\rho)}\;.
\]
Hence, setting $\omega_{min}\sim v_F/x_0$, where $x_0$ is
the mean distance between scatterers,
we find the following condition on doping
\be
x_0 \gg \frac{1}{k_F} \left(\frac{D}{V_0}\right)^{\displaystyle
2/(1-K_\rho)}\;.
\label{DcondA}
\ee
Notice also that a {\em short-range} character of
the impurity scattering was implied:
$a_0 \sim 1/k_F$, where $a_0$ is the radius of
an individual scattering potential.
Otherwise, we should impose an additional condition:
\be
x_0 \gg a_0\;.
\label{DcondB}
\ee

Thus, if the conditions (\ref{DcondA}) --
low doping -- and (\ref{DcondB}) -- short-range
scattering -- are satisfied, we arrive at a novel
physical picture for doped quasi-1D
materials: the impurity potential renormalizes
to large values thereby breaking up the
electron system as shown in the Fig.\ref{quasi}; the electron
tunneling across weak links and
from chain to chain
can then be treated by perturbation theory.
This picture essentially differs from those previously discussed in the
the literature, which deal with perturbation expansions
in the disorder potential\cite{1d,GS}.

Let us briefly outline some consequences.
\begin{itemize}
\item[{\bf (i)}] The Friedel oscillation
(\ref{friedelosc}) will be frozen in, depending upon
the concrete realization of doping. Of course,
this oscillation will be smeared
out by impurity averaging.  Still, an effect should remain
in the density-density correlation
function which acquires an extra {\em doping dependent}
$2k_F$ component steming from the impurity
averaged Friedel oscillation. This should
in principle be experimentally measurable (e.g., by neutron scattering).
\item[{\bf (ii)}] The calculation of
conductivity is, under these circumstances, an interesting but puzzling task
dealing with
a type of random resistance network.
\item[{\bf (iii)}] At low temperatures the electron tunneling
between chains becomes relevant, so that
the 3D character of the problem takes over and
a broken symmetry ground state forms
(typically charge-density wave)\cite{1d,Bourbonnais,Brazovskii};
the effect of small doping (which increases a phase
disorder but stabilizes the Friedel oscillation)
on the formation of this state is unclear.
\end{itemize}
These issues deserve, in our opinion, further investigations.

\subsection{Green's function}

The electron Green function is given by ($t>0$):
\be
G_{ss'}(x,y;t)=-i
\langle \psi_{s}(x,t)\psi_{s'}^\dagger(y,0)\rangle
= \sum_{a,b=\pm 1}a b {\rm e}^{ik_F(ax-by)}G_{ss'}^{R}(ax,by;t)\;,
\label{greendec}
\ee
where we have substituted Eq.(\ref{decomposition}) and defined
\[
G_{ss'}^{R}(x,y;t)=-i
\langle \psi_{Rs}(x,t)\psi_{Rs}^\dagger(y,0)\rangle \;.
\]
Making use of Eq.(\ref{fermiopfinal}), we find
\bea
&~&
G_{ss'}^{R}(x,y;t)=-
\frac{i\delta_{ss'}}{2\pi\alpha}
\prod_\nu \left\{
[P(2x)P(2y)]^{s_\nu c_\nu/2}
\left[P(x^{(-)}- v_\nu t)\right]^{c_\nu^2 /2}
\left[P(x^{(-)}+v_\nu t)\right]^{s_\nu^2 /2}
\right. \nonumber\\&~&\left.
\left[P(x^{(+)}-v_\nu t)P(x^{(+)}+v_\nu t)\right]^{-s_\nu c_\nu /2}
\right\}
{\rm e}^{i\Phi(x,y,t)}
\;,
\label{green}
\eea
where $x^{(\pm)}=x\pm y$, and the phase factor is given by:
\begin{eqnarray*}
\Phi(x,y,t)&=& \frac{\pi}{L} x^{(-)}
-\frac{\pi}{4L}\sum_\nu v_{\nu N} t
+ \frac{1}{2} \sum_\nu \left[ c_\nu^2 f(x^{(-)}-v_\nu t) \right. \\
&-& \left. s_\nu^2 f(x^{(-)}+v_\nu t) + c_\nu s_\nu f(x^{(+)}+v_\nu t)
- c_\nu s_\nu f(x^{(+)}-v_\nu t)\right]
\end{eqnarray*}

It is straightforward to calculate the finite temperature version of
Eq.(\ref{green}), that is
however not of an immediate interest.

Let us now discuss some limiting cases of the expression (\ref{green}).
\begin{itemize}
\item[{\bf (i)}] It is interesting to understand how the ``bulk" behavior of
Green's
function is recovered. Clearly, one should impose the condition
that the relative distance between points
$x$ and $y$ is much less than all the distances to the boundaries:
\be
x^{(-)} \ll \min \{ x,y,L-x,L-y \} \;.
\label{condd}
\ee
However, even if the above condition if fulfilled, the boundaries  still
influence the Green function behavior provided that the time interval $t$
between the creation and the annihilation of the extra electron
is large enough to allow the excitations to reach one of the boundaries and
to be reflected from it. Hence the additional condition:
\be
(v_\rho t, v_\sigma t)  \ll \min \{ x,y,L-x,L-y \} \;.
\label{condt}
\ee
Provided the conditions (\ref{condd}) and (\ref{condt}) are satisfied,
Green's function (\ref{green}) takes the following asymptotic form:
\be
G_{ss'}^{R}(x,y;t)=-
\frac{i\delta_{ss'}}{2\pi\alpha}
\prod_\nu \left\{
\left[ \frac{\alpha}{x^{(-)}-v_\nu t + i\alpha}\right]^{c_\nu^2 /2}
\left[ \frac{\alpha}{x^{(-)}+v_\nu t -i \alpha}\right]^{s_\nu^2 /2}
\right\}\;,
\label{greenbulk}
\ee
which is the electron Green function of the translation invariant system
$(t>0)$.
{}From the bulk behavior of Green's function (\ref{greenbulk})
one can deduce, for instance, that
\be
\langle \psi_{s}(x,t)\psi_{s}^\dagger(x,0)\rangle \sim
\left(\frac{1}{t}\right)^{\frac{1}{2}\sum_\nu\left(K_\nu+K^{-1}_\nu\right)}\;.
\label{psidimbulk}
\ee
\item[{\bf (ii)}] On the other hand, if our $\psi$-operator is close to the
boundary, the condition (\ref{condd}) can not be satisfied any more and one
should use the general formula (\ref{green}). Consider the case $x=y=0$. Of
course, strictly at the boundary, $x=0$, the electron operator vanishes,
Eq.(\ref{boundcond}). So, by writing $x=0$ we mean that $x$ is of the order
of $\alpha$. In this case we find:
\be
G_{ss'}(0,0;t)\sim\delta_{ss'}
\left(\frac{1}{t}\right)^{\sum_\nu K^{-1}_\nu/2}\;.
\label{psidimbound}
\ee
The latter formula holds actually for any $x$ and $y$ satisfying
\be
(x,y)\ll(v_\rho t, v_\sigma t)
\label{condboundop}
\ee
and means that the power law exponent of the Green function
at the boundary differs from the bulk one. This result is
implicit in the Kane and Fisher
treatment\cite{K&F,Glazman} and also agrees with conformal field theory
considerations\cite{Aff}.
\item[{\bf (iii)}] Different exponents come into play in the {\em static
limit}:
$G(x,y;t=0)$. For $t=0$ and $x=y$ the expression (\ref{green}) coincides,
of course, with (\ref{friedelosc}). In the limit $x=0$, $y\gg\alpha$ (but $y\ll
L$),
however, one finds:
\be
G(0,y;0)\sim \left(\frac{1}{y}\right)^{\displaystyle
(1/8)\sum_\nu \left( K_\nu + 3/K_\nu \right)} \; .
\label{dist}
\ee
Notice the difference between (\ref{psidimbound}) and (\ref{dist}).
\end{itemize}

\subsection{Ward's identity}

Because of the novelty of the bosonization
approach to open boundary systems,
the question
arises whether any important physical processes
have been missed in the above consideration.
Put another way, is it possible to derive the
Green function (\ref{green}), and other correlation
functions, by means of an alternative method
not relying on specific tricks of the bosonization
approach [like Eq.(\ref{fermiop}) to Eq.(\ref{fermiopfinal})]?
Alternative method of deriving the correlation functions
of 1D interacting Fermi systems, based on the
classification of diagrams (Ward's identity),
have been devised by Dzyaloshinskii and Larkin\cite{DL}.
Everts and Schulz realized that
Ward's identity can equivalently be derived using equations of motion
for electron density operators\cite{ES}.

The starting point, as in Section II, is to formulate
the problem entirely in terms of right
moving fields. The full Hamiltonian then reads:
\be
H=\int_{-L}^{L} dx \left\{ v_F \psi^{\dagger}_{R} (x)(-i\partial_x )
\psi^{\phantom{\dagger}}_{R} (x)
+\frac{g}{2}\rho_R(x)\rho_R(x)+
\frac{\tilde{g}}{2}\rho_R(x)\rho_R(-x) \right\}\;.
\label{Wham}
\ee
(For the sake of clarity we consider the simplest case of spinless fermions.)

The Dyson equation for the single-particle Green function,
\[
G^R_{pp'}(t)=-i\langle T\{ c_p(t) c_{p'}^\dagger (0) \} \rangle\;,
\]
where the operator $c_p$ is defined by Eq.(\ref{Rff}) with $k_F+p\to p$,
takes the form:
\be
\left( i \partial_t -v_Fp \right) G^R_{pp'}(t)= \delta_{pp'}\delta(t)
-\frac{1}{2L} \sum_q \left[ g F^q_{p-q,p'}(t,t)+
\tilde{g} F^{-q}_{p-q,p'} (t,t) \right] \;,
\label{WDyson}
\ee
being $F$ the two-particle vertex function defined by
\be
F^q_{p,p'}(t,t')= -i\langle T\{\rho_q(t)c_p(t')c_{p'}^\dagger(0)\} \rangle
\label{WF}
\ee
with
\[
\rho_q=\sum_p c_p^\dagger c_{p+q}\;.
\]

The crucial step in solving the 1D problem
is to find a relation between
the vertex function $F$ and the Green function
$G$ which enables one to express the former in
terms of the latter thereby resulting in a
closed equation for $G$. Since
the non-local interaction
in (\ref{Wham}) does not conserve momentum,
the diagrammatic approach of Ref.[$\!\!$\onlinecite{DL}]
is, in our case, less convenient than the
equations of motion method of Ref.[$\!\!$\onlinecite{ES}].

Standard commutation relations\cite{ML}
\[
\left[ \rho_{q}, \rho_{-q'} \right] =
\frac{qL}{\pi} \delta_{qq'}\; ; \;\;\;\; q,q'>0
\]
(the $1/2$ difference with Ref.[$\!\!$\onlinecite{ML}]
is due to $q$ taking the values $\pi n/L$)
lead to the following equation of
motion for the density operator:
\be
\left( i\partial_t - v_F q \right) \rho_q =
\frac{g}{2\pi} q \rho_q + \frac{\tilde{g}}{2\pi} q \rho_{-q}
\label{Weqmot} \;.
\ee

Deriving Eq.(\ref{WF}) in $t$ and substituting Eq.(\ref{Weqmot}),
we arrive at the version of Ward's
identity applicable to the present problem:
\be
\begin{array}{c}
\left( i\partial_t - v_F q \right) F^q_{p,p'}(t,t') = \frac{\displaystyle
g}{\displaystyle
2\pi} q F^q_{p,p'}(t,t')
+ \frac{\displaystyle
\tilde{g}}{\displaystyle
2\pi} q F^{-q}_{p,p'}(t,t')\\
+i\delta (t-t') G^R_{p+q,p'}(t) - i \delta (t) G^R_{p,p'-q}(t'-t)\;.
\end{array}
\label{WWi}
\ee

The Ward identity (\ref{WWi}) can easily be solved
with respect to the vertex function $F$
(in $\omega$-space). Substituting this solution into
Eq.(\ref{WDyson}) and transforming back
to the $(x,t)$-space, we find, after algebraic
manipulations, the following closed equation for the
Green function ($t>0$):
\be
\left( \partial_t + v_F \partial_x \right)
\ln G^R(x,y,t) = K(x,y,t)\;,
\label{Wgeeneq}
\ee
with
\be
K(x,y,t)= \frac{(v_F-v)sc}{x+y-vt} +
\frac{(v_F+v)sc}{x+y+vt} -\frac{v_Fsc}{x}
- \frac{(v_F-v)c^2}{x-y-vt} - \frac{(v_F+v)s^2}{x-y+vt}\;.
\label{Wkernel}
\ee
The limit $L\to\infty$ (semi-infinite system) is taken.
Here and below we omit (imaginary) short-time
cut-off terms (which can easily be restored).
Analogously to the case of fermions with spin
(see Section II), we have defined
\[
c=\cosh (\varphi), \; s=\sinh (\varphi ) ; \;\;
\tanh (2\varphi)=-\frac{\tilde{g}}{2\pi v_F + g}
\]
and the renormalized sound velocity $v$ is given by
\[
v=\sqrt{\displaystyle
\left[v_F+\left(\frac{g}{2\pi}\right)\right]^2-
\left(\frac{\tilde{g}}{2\pi}\right)^2}\;.
\]

Among many solutions of Eq.(\ref{Wgeeneq}),
the one which coincides with the free fermions Green function
for the non-interacting case and is correctly
symmetrized with respect to $x$ and $y$, can be written
in the form:
\be
G^R (x,y,t)=-\frac{i}{2\pi\alpha}
\left[\frac{(x+y-vt)(x+y+vt)}{4xy}\right]^{sc}
\left[\frac{\alpha}{x-y-vt}\right]^{c^2}
\left[\frac{\alpha}{x-y+vt}\right]^{s^2}\;.
\label{Wgreen}
\ee

The Green function (\ref{Wgreen}) coincides with
the appropriate asymptotic form of spinless version
of Eq.(\ref{green}) thus demonstrating the
equivalence of ``purely" bosonization methods of
Section II and the approach based on Ward's
identity for Fermi systems with open boundaries.
[We remind that the total Green function $G(x,y,t)$
is determined by $G^R (x,y,t)$ via the relation
(\ref{greendec}).]

The Ward's identity method can straightforwardly
be extended to the case of finite $L$, finite
temperatures, as well as to the case of fermions
with spin. We are not presenting these calculations
here since they anyway lead to the results identical
to those of the bosonization technique.

\section{Tunneling at the boundaries}
\label{sec:tunn}

In this section we study several tunneling processes at the
boundary of the wire which might be relevant for experimental applications.

\protect \subsection{Two terminal conductance}

Consider a semi-infinite wire coupled to a normal (3D) metal through
an insulating barrier, as shown in Fig.\ref{contact}. A potential
difference $V$ is applied between the metal and the wire and we assume that
the potential drop occurs just across the barrier, whereas both the
metal and the wire have homogeneous chemical potentials. The Hamiltonian
describing the system can be written in the form:
\begin{equation}
\hat{H}=\hat{H}_M + \hat{H}_W + \hat{T} - V \hat{Q}
\label{ham-VQ}
\end{equation}
where $\hat{H}_{M(W)}$ is the Hamiltonian of the isolated metal(wire),
$\hat{T}$ describes the tunneling between them:
\[
\hat{T}= \sum_s \int_W dx \int_M d \vec{r} \; T(x,\vec{r})
\left[ \psi_{Ws}^\dagger(x)\psi_{Ms}^{\phantom{\dagger}}(\vec{r})
+ \;{\rm H.c.}\right]
\]
where $T(x,\vec{r})$ the tunneling matrix element between the
wire and the metal, which we take limited to
the vicinity of the barrier. The last term in (\ref{ham-VQ}) is responsible
for the potential difference between the wire and the metal, $\hat{Q}$ being
their charge difference
\[
\hat{Q}=e(N_W-N_M),
\]
where $N_{M(W)}$ is the total number of electrons in the
metal(wire).

The potential drop induces a current $\hat{I}$ which is defined by:
\be
\hat{I}(t)=\frac{d\hat{Q}}{dt}=-2ie \sum_s \int_W dx \int_M d\vec{r}
\; T(x,\vec{r})
\left[ \psi_{Ws}^\dagger(x)\psi_{Ms}^{\phantom{\dagger}}(\vec{r})
- {\rm H.c.}\right].
\label{t-c}
\ee
In linear response the average current at time $t$ is given by
\[
\langle \hat{I}(t) \rangle =
i\int dt' \theta(t-t') \langle [ \hat{I}(t),\hat{Q}(t') ] \rangle
V(t')
\]
Since $[ \hat{I},\hat{Q} ] = 4ie^2 \hat{T}$, and $\hat{Q}$ is conserved
in the absence of tunneling, the above correlation  function vanishes
if the thermal average is taken over the eigen-states of the
Hamiltonian (\ref{ham-VQ}) with $\hat{T}=0$. Therefore one needs
the next order correction in the tunneling, which leads to the
following expression
\be
\langle \hat{I}(t) \rangle =
i \int dt' \theta(t-t') \int dt'' \theta(t'-t'')
\langle [ \hat{I}(t),\hat{I}(t'') ] \rangle_0
V(t')
\label{current-VQ}
\ee
where the thermal average $\langle ... \rangle_0$ is now taken
at $\hat{T}=0$.

Due to the relation (\ref{t-c}), the current-current correlation function
reduces to a product of the local single particle Green functions for
the metal and the wire. As we have shown previously the local wire Green
function behaves as
\[
\langle \psi_{Ws}^{\phantom{\dagger}}(0,t)\psi_{Ws}^\dagger(0,0)
\rangle \sim \left( \frac{1}{t} \right)^{\sum_\nu K_\nu^{-1}/2}
\]
while the metal Green function is
\[
\langle \psi_{Ms}^{\phantom{\dagger}}(0,t)\psi_{Ms}^\dagger(0,0)
\rangle \sim \left( \frac{1}{t} \right).
\]
The latter result is true independently of the presence of
electron-electron interaction in the 3D metal, as
follows from the quasi-particle pole residue $Z$ being finite in the
Fermi liquid theory.

Plugging these results in (\ref{current-VQ}) we find for
low frequency conductance
\[
G(\omega) = \frac{I(\omega)}{V(\omega)}= G_0
\left( \frac{\omega}{D} \right)^{{\sum_\nu K_\nu^{-1}/2} -1}
\]
where $G_0$ is a frequency independent constant which is determined by
details of the barrier and $D$ is a high energy cut-off related
to the metal and wire bandwidths. At zero frequency the temperature
dependence of the conductance is given by the same formula with
the temperature replacing the frequency.

This result enables us to discuss the more realistic geometry
of Fig.\ref{contact-b}. In this case the voltage drop $V$ occurs
at the two insulating barriers, i.e. $V=V_1+V_2$ (the indices
refer to each barrier), while the
currents flowing across each barrier are the same $I=I_1=I_2$.
The latter condition holds since no accumulation of charge
can occur in the wire. Therefore the two terminal conductance
\[
G=\frac{G_1 G_2}{G_1+G_2} \propto
\left( \frac{\omega}{D} \right)^{{\sum_\nu K_\nu^{-1}/2} -1}
\]

Thus, in the physically relevant situation of
spin isotropic repulsive interaction ($K_\rho<1$ and $K_\sigma=1$),
the two terminal conductance vanishes at low frequency (low temperature)
with a power law. The conductance of a Luttinger liquid with
an impurity has been predicted to vanish by Kane and Fisher\cite{K&F}.
Here we have shown that such an effect should occur even in the case
of a perfectly {\em clean} wire. This reflects the irrelevance of the tunneling
$\hat{T}$ between the wire and the normal metal as a consequence of the
electron-electron interaction in the wire. Thus a two terminal
experiment as the one we discussed above can provide a simple way of
probing the Luttinger liquid behavior in quantum wires.
Notice however that those wires should be sufficiently long since
the dimensional quantization introduces a low energy cut-off
$v_F/L$ below which the conductance becomes temperature independent
and moreover gives rise to a finite charging energy
(causing Coulomb blockade type phenomena)
which we have neglected.

\protect\subsection{Boundary operators}

Similarly to what we have shown for a
normal metal - quantum wire - normal metal
contact, one can imagine other experimental set ups in which different
boundary processes are involved. In the case we just studied
the corresponding boundary operator was the single Fermi field operator.
For example, at the contact between the quantum wire and a
standard 3D superconductor the tunneling of electron pairs should be
considered\cite{NoteH}.
In Table \ref{table}, worked out by applying Eq.(\ref{fermiopfinal}),
we give a list of possible boundary operators
$\hat{O}_b$ with their scaling dimensions $x_b$ defined by
\[
\langle \hat{O}_b(t) \hat{O}_b^\dagger(0)\rangle \propto \left(
\frac{1}{t} \right)^{2x_b}.
\]

\section{Magnetic impurity}
\label{sec:kondo}
In this section we consider the effects of a magnetic impurity
in an interacting wire. The coupling to the conduction electrons
is provided by an antiferromagnetic exchange $J$ and a local potential
$V$. On general grounds one expects $V\gg J$, but even in the
unphysical case $V=0$, a local potential will be generated by the
exchange coupling\cite{boss}.
The $V=0$ problem has been previously studied by Lee and Toner\cite{Lee}
and by Furusaki and Nagaosa\cite{Kondo}.
In Ref.[$\!\!$\onlinecite{Kondo}] it was found that for repulsive
electron-electron interaction the low temperature fixed point corresponds to
the screened impurity spin, similarly to what happens in the conventional
single channel Kondo problem for non--interacting electrons.

Here we study the opposite limit $V\gg J$. In this case it is more
appropriate to first diagonalize the Hamiltonian
with only the local potential. For repulsive interaction that corresponds
to cutting the wire at the impurity site and treating the
residual tunneling through the barrier as a (irrelevant)
perturbation\cite{K&F}. Taking into account also the spin exchange,
one can in general write the following impurity Hamiltonian:
\begin{eqnarray}
H_{imp}&=& J\vec{S}\sum_{i=1,2}\sum_{\alpha\beta}
\psi^\dagger_{i\alpha} \vec{\sigma}_{\alpha\beta}
\psi^{\phantom{\dagger}}_{i\beta} +
\Gamma \vec{S}\sum_{\alpha\beta} \left(
\psi^\dagger_{1\alpha} \vec{\sigma}_{\alpha\beta}
\psi^{\phantom{\dagger}}_{2\beta} + {\rm H.c.}\right)\nonumber\\
&+& t\sum_\alpha\left( \psi^\dagger_{1\alpha}
\psi^{\phantom{\dagger}}_{2\alpha} + {\rm H.c.}\right),
\label{Hamimp-Kondo}
\end{eqnarray}
where 1(2) corresponds to the right(left) side of the impurity,
$\vec{S}$ is the impurity spin-1/2 operator and $\sigma^a$ are
the spin-1/2 matrices. The first term represents the exchange
interaction of the electrons of each lead with the impurity spin.
The second one corresponds to tunneling processes with spin flip,
and the third one to the tunneling without spin flip.

For $\Gamma=t=0$ the model is equivalent to the two channel Kondo
model, which is known to exhibit anomalous behavior.
Finite $\Gamma$ and $t$ will introduce anisotropy between the two
channels. For non-interacting electrons this anisotropy is
known to be a relevant perturbation which brings the system back to
the single channel behavior at low temperature\cite{classic}.
A repulsive interaction ($K_\rho<1$, assuming $K_\sigma=1$
to assure spin isotropy) changes the bare dimension of $\Gamma$
getting it irrelevant, without affecting $J$ (which remains marginal).
The resulting renormalization group equations for weak interaction are:
\begin{eqnarray*}
\frac{d J}{d\ln L} &=& \frac{1}{2\pi v_F} (J^2+\Gamma^2),\\
\frac{d \Gamma}{d\ln L} &=& \left(1-\frac{1}{K_\rho}\right)\Gamma
+ \frac{1}{\pi v_F} \Gamma J .
%\frac{dt}{d\ln L} &=& \frac{1}{2}\left(1-\frac{1}{K_\rho}\right) t.
\end{eqnarray*}
Although a small $\Gamma$ will firstly decrease under scaling due to
$K_\rho<1$, the contemporary increase of $J$ will ultimately
drive $\Gamma$ to larger values. Therefore the relevance of
$\Gamma$ should be more appropriately studied for large values of
$J$. The best way to proceed is then to first assume $\Gamma=0$,
let $J$ flow to the strong coupling fixed point, and analyse
around it the relevance of a small $\Gamma$.

We bosonize the Fermi fields according to the procedure outlined in
section II. We introduce charge and spin bosonic variables for
each lead $\phi_{1\nu}$ and $\phi_{2\nu}$ ($\nu=\rho,\sigma$),
and their symmetric and antisymmetric
combinations $\phi_{s\nu}$ and $\phi_{a\nu}$. In terms of these
fields the Hamiltonian (\ref{Hamimp-Kondo}) reads
\begin{eqnarray}
H_{imp} &=& \frac{J_\perp}{2\pi\alpha}\left[
S^+ {\rm e}^{i\phi_{s\sigma}} \cos(\phi_{a\sigma}) + {\rm H.c.}
\right] + \frac{J_z}{2\pi}S^z\partial_x \phi_{s\sigma}
+ \frac{\Gamma_\perp}{2\pi\alpha}\left[
S^+ {\rm e}^{i\phi_{s\sigma}}
\cos\left(\frac{\phi_{a\rho}}{\sqrt{K_\rho}}\right) + {\rm H.c.}
\right]\nonumber\\
&+& \frac{\Gamma_z}{\pi\alpha} S^z\sin(\phi_{a\sigma})
\sin\left(\frac{\phi_{a\rho}}{\sqrt{K_\rho}}\right)
+ \frac{2t}{\pi\alpha}\cos(\phi_{a\sigma})
\cos\left(\frac{\phi_{a\rho}}{\sqrt{K_\rho}}\right).
\label{Hamimp-Kondo-bos}
\end{eqnarray}
In the spirit of the Emery-Kivelson solution to
the two-channel Kondo model\cite{E&K},
we can get rid of the phase factors
involving $\phi_{s\sigma}$ by performing the canonical transformation
\[
U={\rm e}^{-iS_z\phi_{s\sigma}}.
\]
The transformed Hamiltonian is simply
\begin{eqnarray}
H_{imp} &=& \frac{1}{\pi\alpha}
S^x\left[J_\perp \cos(\phi_{a\sigma})
+\Gamma_\perp \cos\left(\frac{\phi_{a\rho}}{\sqrt{K_\rho}}\right)
\right]
+ \frac{\lambda}{2\pi}S^z\partial_x \phi_{s\sigma}
\label{Hamimp-Kondo-bos-trans}\\
&+& \frac{\Gamma_z}{\pi\alpha} S^z\sin(\phi_{a\sigma})
\sin\left(\frac{\phi_{a\rho}}{\sqrt{K_\rho}}\right)
+ \frac{2t}{\pi\alpha}\cos(\phi_{a\sigma})
\cos\left(\frac{\phi_{a\rho}}{\sqrt{K_\rho}}\right). \nonumber
\end{eqnarray}
where $\lambda=J_z-2\pi v_F$. If $\lambda=\Gamma_z=t=0$ the
Hamiltonian (\ref{Hamimp-Kondo-bos-trans}) can be easily
studied since $S^x$ commutes with the Hamiltonian and can be given
a definite value $\pm 1/2$. For a fixed $S^x$ the Hamiltonian
is equivalent to the one describing impurity scattering in a Luttinger liquid
(for each field $\phi_{a\sigma}$ and $\phi_{a\rho}$).
It is known that a local $\cos (\beta \phi)$-term is relevant if
$\beta<\sqrt{2}$, marginal if $\beta=\sqrt{2}$, and irrelevant otherwise.
Therefore the $J_\perp$ operator is always relevant, whereas
$\Gamma_\perp$ is relevant only if $1/2 < K_\rho <1$.

If both $J_\perp$ and $\Gamma_\perp$ are relevant, the low temperature
fixed point behavior is that of the single channel Kondo model, as
we previously discussed. Just this case has been analysed in detail
by Furusaki and Nagaosa\cite{Kondo}. Thus, we have shown that the presence
of a strong potential scattering does not modify their conclusions,
provided that $K_\rho\ge 1/2$.

The new feature which arises from our analysis is that for $K_\rho<1/2$ the
two channel Kondo behavior is stable with respect to a channel
asymmetry. The behavior of the model around the two channel fixed
point can be performed in a way similar to that of
Ref.[$\!\!$\onlinecite{E&K}].
Without repeating the calculation we just remind that the low temperature
impurity susceptibility $\chi\sim \ln (1/T)$ and the specific heat
$C_V\sim T\ln(1/T)$. This situation could be realized in quantum wires
with long-range Coulomb interaction, which may result in a very small
$K_\rho$, see Ref.[$\!\!$\onlinecite{Schulz}].

\section*{Acknowledgments}

This work is already old. In the course of time,
we have benefited from discussions with many our colleagues.
We are especially thankful to Arkadii Aronov,
Matthew P.A. Fisher, and Philippe Nozi\`{e}res for correspondence and for
numerous helpful and encouraging discussions.

\appendix
\section{}
\label{app-A}

The computation of average values of exponentials containing phase fields
requires the knowledge of the sum:
\be
S(z)=
\sum_{q>0}\frac{\pi}{qL}{\rm e}^{iqz-\alpha q}
=\ln\left(\frac{L}{\pi\alpha}\right) + \ln P(z) + if(z)
\ee
where :
\[
P(z)=  \frac{\pi \alpha}{ 2L\sqrt{
\sinh ^2  \left( \frac{\pi \alpha}{2L} \right) +
\sin ^2  \left( \frac{\pi z}{2L} \right) } };
\]
\[
f(z)= \tan^{-1} \left[ \frac{
\sin \left( \frac{\pi z}{L} \right) }
{ {\rm e}^{\pi \alpha/L} - \cos\left( \frac{\pi z}{L} \right) }
\right].
\]
Let us consider the limit of these functions for $z$ close
to one of the boundaries. For $|z|\ll L$ (but $|z|\gg\alpha$) we have
\[
P(z)= \frac{\alpha}{|z|} ;
\]
\[
f(z)=\tan^{-1}\left(\frac{z}{\alpha}\right)\to \frac{\pi}{2} {\rm sign}(z).
\]
The same expression holds at the other boundary
($z\to 2L$) with the replacement $z\to z-2L$.
\section{}
\label{app-B}

As an interesting application of the open boundary bosonization
we consider the following problem: let us imagine that
in a spinless Luttinger liquid
on a ring of length $L$ we insert an impurity at the origin.
This impurity has two possible
states, which can be thought as two spin states $\mid \uparrow
\!\!(\downarrow)\rangle$.
We model the impurity as a local back--scattering potential\cite{K&F}, and
we assume that the sign of this
potential depends on the impurity spin. This amounts to add
to the electron Hamiltonian a term of the form
\be
V\sigma_z \cos(\phi_R + \phi_L).
\ee
where the Pauli matrix acts on the impurity states, and
$\phi_{R(L)}$ is the phase field corresponding to the
right(left) moving electrons. Since there is no term in the Hamiltonian
which flips the impurity spin, $\sigma_z$ can be given a fixed
value, which does not evolve with time.
If the interaction is repulsive (parametrized by the exponent
$K<1$) we know from Ref.[$\!\!$\onlinecite{K&F}] that the impurity at low
energy
effectively acts as an infinite barrier, thus cutting the chain
into two disconnected leads of length $L$.
Consider now the following x--ray edge type of problem:
at time $\tau<0$ the impurity is in the state $\mid \downarrow\rangle$.
Suddenly at $\tau=0$ the spin is reversed, thus changing sign of the
scattering potential, and finally at time $\tau=t$ the spin is reversed back.
This process is described by the correlation function:
\begin{equation}
\chi(t)=\langle \downarrow \mid
\sigma^-(t) \sigma^+(0) \mid \downarrow \rangle\;.
\label{corr-appB}
\end{equation}
Can we calculate the long time behavior of (\ref{corr-appB}) ?
We recently encountered a similar problem analysing the
four channel Kondo model\cite{our-four}, and we think it may be relevant
also for other impurity models.
The way to evaluate (\ref{corr-appB}) is similar to the standard
bosonization approach to the x--ray edge singularity. Notice that
the unitary operator
\[
U=\exp{\left[ i\frac{\pi}{2}J\right]},
\]
where $J=N_R-N_L$ is the total current, has the following property
\[
U \left[H_0 + V\cos(\phi_R + \phi_L)\right] U^\dagger
= H_0 - V\cos(\phi_R + \phi_L),
\]
i.e. it changes the sign of the potential term
without modifying the bulk part of the Hamiltonian $H_0$.
Thus $U$ is equivalent to the spin flip operator. This implies that
the correlation function (\ref{corr-appB}) can also be written as
\[
\chi(t)= \langle \downarrow \mid
U(t) U^\dagger (0)\mid \downarrow \rangle\;,
\]
and in this representation its evaluation is straightforward.
Since the low energy fixed point corresponds to cutting the ring at
the origin, we use the open boundary bosonization to rewrite
the current operator
\[
J=N_R - N_L =
\int_0^L dx \rho_R(x) - \rho_L(x)
= \int_0^L dx \rho_R(x) - \rho_R(-x).
\]
By performing the
Bogolubov rotation to get rid of the bulk interaction we get
\[
J = \frac{1}{\pi \sqrt{K} }(\phi(L) - \phi(0) )
\]
and therefore
\[
U = \exp\left[ \frac{i}{2\sqrt{K}}
(\phi(L) - \phi(0) ) \right]
\]
Since the fields $\phi$ are now free Bose fields with
logarithmic correlation, we immediately see that
the dynamics of the total current is
characterized by the following correlation function
($t\ll L/v_F$):
\[
\langle J(t) J(0)\rangle = \frac{2}{\pi^2 K} \ln t
\]
This result agrees with the analysis of Ref.[$\!\!$\onlinecite{Kolya}].
Hence the correlation function (\ref{corr-appB}) takes the form:
\be
\chi(t) \propto \left( \frac{1}{t} \right) ^{\frac{1}{2K}}\;.
\label{result-appB}
\ee
$\chi(t)$ has the power law behavior typical of the
x--ray edge singularity, but the exponent is determined by the
interaction only and is not bounded from above.

\section{}
\label{app-C}

This Appendix is intended to show that a long-range electron-electron
interaction
(i.e. interaction of finite radius $R$) does not modify the conclusions
of Section II; namely, that the exponents of correlation functions are
determined
by zero-momentum Fourier components of interaction constants.
In the case of long-range interaction the bosonic Hamiltonian
(\ref{ham-nondiag}) is
no more diagonal in $q$-space. We proceed in two steps: firts, we demonstrate
how the diagonal part
of the interaction can be separated from the off-diagonal one and, second, we
observe that the latter
is irrelevant (does not contribute to the exponents).

The interaction term (\ref{gint}) is now of the form
\be
\frac{1}{2}\int_{0}^{L}dx \int_{0}^{L}dy g(x-y) \left[
\rho_{\rho(\sigma) R} (x)\rho_{\rho(\sigma) R} (y)
+\rho_{\rho(\sigma) L} (x)\rho_{\rho(\sigma) L} (y)\right]
\label{Ngint}
\ee
and the $\tilde{g}$ term (\ref{gtildeint}) changes analogously. (The index
$\nu=\rho,\sigma$
is suppressed but the results are valid for both charge and spin
sectors.)

Simplifying the problem (but not affecting qualitative results) we work in the
limit
of semi-infinite system ($L\to\infty$). Then, in $q$-space, the Hamiltonian
takes the form:
\be
H=H_0+H_{int}
\label{Nham}
\ee
with the continuum version of the free part,
\[
H_0= \int_0^\infty dq  v_F q b^\dagger_q b^{\phantom{\dagger}}_q \;,
\]
and the interaction term:
\bea
H_{int}=-\frac{1}{4\pi}\int_0^\infty dq_1\int_0^\infty dq_2 \sqrt{q_1 q_2}
&~&\!\!\!\!\!\!\left[
I(q_1,q_2) \left( b^{\phantom{\dagger}}_{q_1} b^{\phantom{\dagger}}_{q_2} +
{\rm H.c.}
\right) \right.\nonumber \\&+& \left. I(q_1, -q_2)
\left( b^{\dagger}_{q_1} b^{\phantom{\dagger}}_{q_2} + {\rm H.c.}
\right) \right]\;,
\label{Nint}
\eea
where
\[
I(q_1,q_2)=g(q_1,q_2)+\tilde{g}(q_1,-q_2)
\]
and $g(q_1,q_2)$ is defined by
\be
g(q_1,q_2)={\rm Re}\frac{1}{\pi} \int_{-\infty}^\infty \frac{dp}{2\pi}
\frac{ \displaystyle
g(p)}{ \displaystyle
(p+q_1+i\delta)(p-q_2-i\delta)}\;.
\label{Ngintegral}
\ee
Here $g(p)$ is the Fourier integral transform of $g(x)$ and $\delta$ a positive
infinitesimal.
The formula (\ref{Ngintegral}) with $\tilde{g}(p)$ instead of $g(p)$ defines
$\tilde{g}(q_1,q_2)$.

Shifting the integration contour in (\ref{Ngintegral}) up in the complex
$p$-plane, we can write:
\be
I(q_1,q_2)=g(q_2)\delta (q_2+q_1) + \tilde{g}(q_2)\delta(q_2 -q_1) + \delta
I(q_1,q_2) \;.
\label{Nseparation}
\ee
The first two terms in (\ref{Nseparation}) stem from the residual of the pole
$p=q_2+i\delta$. They are
responsible for the diagonal part of the interaction. The off-diagonal part,
$\delta I(q_1,q_2)$,
is due to singularities of the functions $g(p)$ and $\tilde{g}(p)$ in the
complex plane
(see Fig.\ref{contour}).
Since the only regular complex function is a constant, the only case when
$\delta I=0$
corresponds to local (in real space) interactions.

In order to illustrate the properties of the
off-diagonal part of the interaction we consider now a specific example:
\be
g(p)=\tilde{g}(p)= \frac{g_0}{2} \ln  \left(
\frac{p^2+\lambda_a^2}{p^2+\lambda_s^2} \right)
\label{Ncoulomb}
\ee
The interaction (\ref{Ncoulomb}) may model a screened Coulomb interaction in
quantum wires (the cut-offs
$\lambda_s$ and $\lambda_a$ are then related to the inverse screening length
and inverse lattice
spacing respectively). Evaluating the branch cut
integral (Fig.\ref{contour}) we find:
\[
\delta g(q_1,q_2)=\frac{g_0}{\pi(q_1+q_2)}\sum_{i=1,2}\left[
\tan^{-1}\left(\frac{q_i}{\lambda_s}\right)
- \tan^{-1}\left(\frac{q_i}{\lambda_a}\right)\right]\;.
\]
It is important to notice that $\delta I$ tends to a constant in the limit of
small $q$:
\[
\delta I(0,0) = \frac{2g_0}{\pi}\left( \frac{1}{\lambda_s}
+\frac{1}{\lambda_a}\right)\;.
\]
Obviously we can, in general, write:
\be
\delta I(q_1,q_2)= g_0 R,\;\;\;\; {\rm for} \;\; (q_1,q_2)\ll 1/R \;.
\label{NIq}
\ee
Here $g_0$ is the interaction strength and $R$ is related to the inverse
distance to the first singularity
of $g(p)$ from the real axis (i.e. to the radius of the interaction).

We can thus re-arrange the total Hamiltonian (\ref{Nham}) writing it in the
form:
\[
H=\tilde{H}+\delta H \; .
\]
Here $\tilde{H}$ is the continuum version of the Hamiltonian
(\ref{ham-nondiag}), it contains $H_0$
and the diagonal in $q$ part of (\ref{Nint}). $\tilde{H}$ can immediately be
diagonalized
[the exponents are determined by $g(q=0)$ and $\tilde{g}(q=0)$]. The remaining
part, in small-$q$ limit,
takes the form:
\[
\delta H = -\frac{g_0 R}{2\pi} \int_0^\infty dq_1 \int_0^\infty dq_2
\sqrt{q_1 q_2}
\left( b^{\phantom{\dagger}}_{q_1} - b^\dagger_{q_1} \right)
\left( b^{\phantom{\dagger}}_{q_2} - b^\dagger_{q_2} \right) \;.
\]
The irrelevance of $\delta H$ is already clear from its scaling dimension (with
is equal to $2$).
One can also demonstrate this more straightforwardly: since $\delta H$ is
quadratic in Bose
fields, $H$ can be diagonalized. Doing so, we find that the leading corrections
(due to $\delta H$)
to the
correlation function $\langle \phi(x) \phi(y)\rangle$ [which behaves as $\ln
(x-y)$ in the absence
of $\delta H$]  drop as $(R/x)^2$, $(R/y)^2$ at large distances thus do not
modifying the
asymptotic behavior of fermionic correlators.

\section{}
\label{app-D}

Here we study how the Luther-Emery solution is modified by open boundary
conditions.
If we define new Fermi operators:
\[
\Psi_\sigma(x)=\frac{1}{\sqrt{2\pi\alpha}}
{\rm e}^{-i\theta_\sigma} {\rm e}^{i \frac{\pi}{L} x \Delta N_\sigma}
{\rm e}^{i\phi_\sigma(x)}\; ,
\]
where $\theta_\sigma$ is conjugate to $\Delta N_\sigma$,
then exactly at $K_\sigma=1/2$ the spin backscattering term
(\ref{sbbos}) assumes a quadratic form in these operators. The
Hamiltonian describing the spin excitations becomes:
\begin{equation}
H_\sigma = v_\sigma \int_{-L}^{L} dx
\Psi^{\dagger}_\sigma (x)(-i\partial_x )
\Psi^{\phantom{\dagger}}_\sigma (x)
- i\Delta
\int_{-L}^{L} dx {\rm sign}(x)
\Psi^{\dagger}_\sigma (x)
\Psi^{\phantom{\dagger}}_\sigma (-x) \; ,
 \label{Hsigma}
\ee
where
\be
\Delta=\frac{g_{bs}}{2\pi\alpha}\; .
\label{Delta}
\ee
The ${\rm sign}(x)$ factor in (\ref{Hsigma}) stems from
the phase factor $f(2x)$ in the $\alpha\to 0$ limit, which
is necessary to preserve the particle--hole symmetry
of the model
\[
\Psi_{sR}(x)\to \Psi_{\bar{s}R}^\dagger (-x)\; ,
\]
or equivalently
\[
\phi_\sigma(x)\to - \phi_\sigma(-x)\; .
\]
This symmetry implies that the spectrum should be
symmetric around zero energy.

Being quadratic in the fermion fields, the Hamiltonian
(\ref{Hsigma}) can be brought into diagonal form
\[
H_\sigma= \sum_\epsilon \epsilon C_\epsilon^\dagger
C_\epsilon^{\phantom{\dagger}}\; ,
\]
where the new Fermi operators $C_\epsilon$ are related to the
old ones by the canonical transformation
\[
C_\epsilon^\dagger = \int dx \chi_\epsilon(x) \Psi_\sigma^\dagger(x)\; .
\]
The wave functions $\chi_\epsilon(x)$ satisfy the following
non--local Schr\"{o}dinger equation:
\be
\epsilon \chi_\epsilon(x) = -iv_\sigma \partial_x \chi_\epsilon(x)
- i \Delta {\rm sign}(x) \chi_\epsilon(-x) \; .
\label{Sch-eq}
\ee
Defining the functions $\chi_\epsilon^{(+)}(x) = \chi_\epsilon(x) $
and $\chi_\epsilon^{(-)}(x) = \chi_\epsilon(-x)$ (for $x>0$),
we can conveniently rewrite (\ref{Sch-eq}) as the system of
differential equations:
\be
\begin{array}{l}
\displaystyle
\epsilon \chi_\epsilon^{(+)}(x) = -i v_\sigma \partial_x
\chi_\epsilon^{(+)}(x)
- i \Delta \chi_\epsilon^{(-)}(x) \; , \\
\displaystyle
\epsilon \chi_\epsilon^{(-)}(x) =
+i v_\sigma \partial_x \chi_\epsilon^{(-)}(x)
+ i \Delta \chi_\epsilon^{(+)}(x) \; .
\end{array}
\label{system-eq}
\ee
Those functions obey the following boundary conditions
\be
\chi_\epsilon^{(+)}(0)= \chi_\epsilon^{(-)}(0)\; ,
\label{cond-a}
\ee
\be
\chi_\epsilon^{(+)}(L)= \chi_\epsilon^{(-)}(L)\; ,
\label{cond-b}
\ee
where the latter derives from (\ref{phi-period}).

Each particular solution of (\ref{system-eq}) has the form
$\exp{(\pm i px)}$.
The dispersion relation turns out to be
\be
\epsilon^2 = (v_\sigma p)^2 + \Delta^2\; ,
\label{disp}
\ee
and the allowed values of $p$ derive from imposing the boundary conditions
(\ref{cond-a},\ref{cond-b}), which lead to
\be
\sin(pL)(i\Delta - \epsilon - v_\sigma p)(i\Delta - \epsilon +
v_\sigma p)=0\; .
\label{sb-eigen}
\ee
The solutions are $p=\pi n/L$ with $n$ a positive integer,
which correspond to energies $|\epsilon|>\Delta$,
and $p=\pm i \Delta/v_\sigma$, with energy $\epsilon=0$.
Therefore, similarly to what happens
in the case of an infinite system, the spin backscattering term
opens a gap $\Delta$ in the continuum spectrum. The novel
feature is the existence of the zero energy states, which are bound
states localized at the boundaries. So, in the $L\to\infty$
limit (semi--infinite system), the wave function of the bound state localized
at  $x=0$
takes the form:
\[
\chi_0 (x) = \sqrt{\frac{\Delta}{v_\sigma}} {\rm e}^{-\Delta |x|/v_\sigma }\;.
\]

\begin{figure}
\caption{Single particle spectrum: for the case of open boundaries the
Fermi surface consists of the single point $k_F$.
}
\label{fermi}
\end{figure}
\begin{figure}
\caption{The system with open boundaries can be described either
in terms of
right and left moving Fermi fields (the upper part of the figure)
or in terms of the right moving field only (the lower part); in the
latter case the density-density interaction becomes non-local.
}
\label{right}
\end{figure}
\begin{figure}
\caption{Slightly doped quasi-1D conductor breaks up into
independent segments of the electron gas (described by the open boundary
analysis), these segments are coupled back by weak tunneling processes
of two types: $t_{perp}$ (tunneling between different chains) and
$t_{12}$ (tunneling between neighbouring segments in the same chain).
}
\label{quasi}
\end{figure}
\begin{figure}
\caption{Contact between a quantum wire and a 3D metal, with an applied
potential difference $V$.}
\label{contact}
\end{figure}
\begin{figure}
\caption{Geometry for a two terminal conductance measurement.}
\label{contact-b}
\end{figure}
\begin{figure}
\caption{Impurity spin $S$ in a quantum wire in the presence of
a strong potential scattering.
}
\label{vspin}
\end{figure}
\begin{figure}
\caption{Integration contour in Eq.(\protect \ref{Ngintegral} \protect$\!\!$).
}
\label{contour}
\end{figure}

\begin{table}
\caption{Boundary operators and their scaling dimensions}
\label{table}
\begin{tabular}{|c|c|c|} \hline
Boundary               & bosonized & exponent \\
operator $\hat{O}_b(t)$&   form    &  $x_b$    \\ \hline
$\psi_{Rs}(t)$ & $\exp \{ i\frac{\phi_\rho(t)}{\sqrt{2K_\rho}}
\pm i\frac{\phi_\sigma(t)}{\sqrt{2K_\sigma}}\}$ & $\frac{1}{4}(K_\rho^{-1}
+K_\sigma^{-1})$ \\ \hline
$\psi_{Rs}^\dagger(t)\psi_{Rs}(t)$
& $\partial \phi_\rho(t) +\partial \phi_\sigma(t)$
& 1 \\ \hline
$\psi_{Rs}^\dagger(t)\psi_{R-s}(t)$
& $\exp \{ \mp i\sqrt{\frac{2}{K_\sigma}}\phi_\sigma(t) \}$
& $K_\sigma^{-1}$ \\ \hline
$\psi_{Rs}(t)\psi_{Rs}(t)$
& $\exp \{ i\sqrt{\frac{2}{K_\rho}}\phi_\rho(t) \pm
i\sqrt{\frac{2}{K_\sigma}}\phi_\sigma(t)\}$
& $K_\rho^{-1}+K_\sigma^{-1}$ \\ \hline
$ \psi_{R-s}(t)\psi_{Rs}(t)$
& $\exp \{ i\sqrt{\frac{2}{K_\rho}}\phi_\rho(t)\}$
& $K_\rho^{-1}$ \\ \hline
\end{tabular}
\end{table}


\begin{references}

\bibitem{1d}
J. S\'{o}lyom, Adv. Phys. {\bf 28}, 201
(1979); V.J. Emery, in {\it Highly Conducting One-Dimensional Solids},
Ed., J.T. Devrese, R.P. Evrard, and V.E. Van Doren (Plenum 1979).

\bibitem{Bourbonnais}
See C. Bourbonnais and L.G. Caron, Int. J. Mod. Phys. B {\bf 5}, 1033 (1991)
and references therein.

\bibitem{Goni}
A. R. Go$\tilde{\rm{n}}$i, A. Pinczuk, J. S. Weiner,
J. M. Calleja, B. S. Dennis, L. N. Pfeiffer, and K. W. West, Phys. Rev.
Lett. {\bf 67}, 3298 (1991).

\bibitem{K&F}
C.L. Kane and M.P.A. Fisher, Phys. Rev. Lett. {\bf 68}, 1220 (1992);
Phys. Rev. B {\bf 46}, 15233 (1992), see also
A. Furusaki and N. Nagaosa, Phys. Rev. B {\bf 47}, 4631 (1993).

\bibitem{everythingelse}
For instance, in the paper
K. Moon, H. Yi, C.L. Kane, S.M. Girvin, and M.P.A. Fisher,
Phys. Rev. Lett. {\bf 71}, 4381 (1993)
a Luttinger liquid description of the
tunneling between quantum Hall edge states has been proposed.
This picture has recently been experimentally confirmed by
F.P. Milliken, C.P. Umbach, and  R.A. Webb (unpublished).
Many other (theoretical as well as experimental)
investigations related to the physics of quantum wires
have recently been carried out, they include studies of Fermi edge
singularities, persistent current, effects of long-range Coulomb
and electron-phonon interactions, Bethe-ansatz solutions, etc.
Some further references
may be found in the review paper A.O. Gogolin,
Ann. Phys. Fr. {\bf 19}, 241 (1994).

\bibitem{Aff}
S. Eggert and I. Affleck, Phys. Rev. B {\bf 46}, 10866 (1992);
E. Wong and I. Affleck, Nucl. Phys. B {\bf 417}, 403 (1994).


\bibitem{ML}
D.C. Mattis and E.H. Lieb, J. Math. Phys. {\bf 6}, 304 (1965).

\bibitem{Lutt}
J.M. Luttinger, J. Math. Phys. {\bf 4}, 1154 (1963).

\bibitem{LP}
A. Luther and I. Peshel, Phys. Rev. B {\bf 12}, 3908 (1975).

\bibitem{Matt}
D.C. Mattis, J. Math. Phys. {\bf 15}, 609 (1974).

\bibitem{Mand}
S. Coleman, Phys. Rev. D {\bf 11}, 2088 (1975),
S. Mandelstam, Phys. Rev. D {\bf 11}, 3026 (1975).

\bibitem{Haldane}
F.D.N. Haldane, J. Phys. C {\bf 14}, 2585 (1981).

\bibitem{Ludwig}
A.W.W. Ludwig,
{\em Methods of Conformal Field Theory in Condensed Matter Physics},
Trieste Summer School Lectures, 1992.

%%\bibitem{notemultiple}
% my suggestion
% do we really need this ? It is a little bit unclear for those who do not
% know the problem.
%%In this paper we have adopted a bosonization
%%scheme in which the expressions for
%%Fermi operators, Eqs. (\ref{fermiop})-(\ref{fermiopfinal}),
%%contain {\em single}
%%$\exp (ik_Fx)$ oscillation factor.
%%A more rigorous approach would include
%We would like to make a
%technical but important comment in
%this relation: expressions for $\psi_{s}(x)$ containing
%%{\em multiple} $\exp (imk_Fx)$, $m=1,2,...,$
%%oscillations
%%[see F.D.N. Haldane, Phys. Rev. Lett. {\bf 25}, 1840 (1981)].
%%However it is important to point out that these terms
%%do {\em not} reproduce all the relevant operators which contribute to
%%the correlation functions of spinning fermions.
%%This concerns, for instance, the $4k_F$-term in the
%%density-density correlation function
% in these days, studying the phonons, I realized one
% maybe important thing. Since K_\rho\to 0 for long range,
% the system is unstable to CDW at all fillings. E.g. you can
% construct the operator which would correspond to a CDW instability
% at quarter filling and you see that is relevant, and so on at
% all commensurate fillings. So do we need to cite this case ?
% We could for example cite the U=\infty Hubbard model
% where the 4k_F is finite but the 2k_F is zero (in the sense
% that is multiplied by zero.
%%which is leading for long-range interactions\cite{Schulz}.
%%Our approach consists to first construct each
%%term in the correlation functions
%%using the fermion language and then apply bosonization formulae
%%(\ref{fermiop},\ref{fermiopfinal}).
%%We are thankful to A.A. Nersesyan and H.J. Schulz for discussions on this
%%point.

\bibitem{L&E}
A. Luther and V. J. Emery, Phys. Rev. Lett. {\bf 33}, 589 (1974).

\bibitem{notescaling}
Of course, this scaling argument is
fully applicable to semi-infinite systems only.
For the case of finite systems,
the scaling process is cut-off by dimensional
quantization energy (see Section IIIA),
so that there is a residual
$g_{bs}$ ({\em small} in $1/L$).

\bibitem{GS}
T. Giamarchi and H.J. Schulz, Phys. Rev. B {\bf 37}, 325 (1988).

\bibitem{Brazovskii}
R.A. Klemm and H. Cutfreund, Phys. Rev. B {\bf 14}, 1086 (1974);
S.A. Brazovskii and V.M. Yakovenko, Zh. Eksp. Teor. Fis. {\bf 89}, 2318 (1985)
[Sov. Phys. JETP, {\bf 62}(6), 1340 (1985)].

\bibitem{Glazman}
See also
K.A. Matveev and L.I. Glazman, Phys. Rev. Lett. {\bf 70}, 990 (1993).

\bibitem{DL}
I.E. Dzyaloshinskii and A.I. Larkin, Zh. Eksp. Teor. Fis. {\bf 65}, 411 (1973)
[Sov. Phys. JETP, {\bf 38}, 202 (1974)].

\bibitem{ES}
H.U. Everts and H. Schulz, Sol. St. Commun. {15}, 1413 (1974).

\bibitem{NoteH}
Josephson current through a quantum wire
has recently been discussed, in the absence of a potential scattering at
contacts,
by R. Fazio, F.W.J. Hekking, and A.A. Odintsov (unpublished). We notice that,
taking into account the potential scattering (even if it initially is very
small),
one inevitably ends up with open boundary conditions so that modifications of
the
bosonization method we considered in the present paper come into play.

\bibitem{boss}
P. Nozi\`{e}res, J. Physique {\bf 39}, 1117 (1978).

\bibitem{Lee}
D.-H. Lee and J. Toner, Phys. Rev. Lett. {\bf 69}, 3378 (1992).

\bibitem{Kondo}
A. Furusaki and N. Nagaosa, Phys. Rev. Lett. {\bf 72}, 892 (1994).

\bibitem{classic}
P. Nozi\`eres and A. Blandin, J. Phys. (Paris) {\bf 41}, 193 (1980).

\bibitem{E&K}
V.J. Emery and S. Kivelson, Phys. Rev. B {\bf 47}, 10812 (1992).

\bibitem{Schulz}
H.J. Schulz, Phys. Rev. Lett. {\bf 71}, 1864 (1993).

\bibitem{our-four}
M. Fabrizio and A.O. Gogolin, Phys. Rev. B {\bf 50}, 17732 (1994).

\bibitem{Kolya}
A.O. Gogolin and N.V. Prokof'ev, Phys. Rev. B {\bf 50}, 4921 (1994).

\end{references}
\end{document}